\documentclass[manuscript]{aastex}

\newcommand\rah{\mbox{$^{\mathrm h}$}}%
\newcommand\ram{\mbox{$^{\mathrm m}$}}%

\slugcomment{To appear in AJ}

\shorttitle{The jet termination shock in Pictor A}
\shortauthors{Tingay \& Lenc}

\begin{document}


\title{A high resolution view of the jet termination shock in a hot spot of the nearby radio galaxy Pictor A: implications for X-ray models of radio galaxy hot spots}


\author{S.J. Tingay}
\affil{Department of Imaging and Applied Physics, Curtin University of Technology, Bentley WA, Australia}
\email{s.tingay@curtin.edu.au}

\author{E. Lenc}
\affil{CSIRO Australia Telescope National Facility, P.O. Box 76, Epping, NSW 2121, Australia}

\author{G. Brunetti, M. Bondi}
\affil{INAF - Istituto di Radioastronomia, Via P. Gobetti 101, 40129 Bologna, Italy}



\begin{abstract}
Images made with the Very Long Baseline Array have resolved the region in a nearby ($z=0.035$) radio galaxy, Pictor A, where the relativistic jet that originates at the nucleus terminates in an interaction with the intergalactic medium, a so-called radio galaxy hot spot.  This image provides the highest spatial resolution view of such an object to date, the maximum angular resolution of 23 mas corresponding to a spatial resolution of 16 pc, more than three times better than previous VLBI observations of similar objects.  The north-west Pictor A hot spot is resolved into a complex set of compact components, seen to coincide with the bright part of the hot spot imaged at arcsecond-scale resolution with the VLA.  In addition to a comparison with VLA data, we compare our VLBA results with data from the HST and \emph{Chandra} telescopes, as well as new Spitzer data.  The presence of pc-scale components in the hot spot, identifying regions containing strong shocks in the fluid flow, leads us to explore the suggestion that they represent sites of synchrotron X-ray production, contributing to the integrated X-ray flux of the hot spot, along with X-rays from synchrotron self-Compton scattering.  This scenario provides a natural explanation for the radio morphology of the hot spot and its integrated X-ray emission, leading to very different predictions for the higher energy X-ray spectrum compared to previous studies.  From the sizes of the individual pc-scale components and their angular spread, we estimate that the jet width at the hot spot is in the range 70 - 700 pc, which is comparable to similar estimates in PKS 2153$-$69 ($z=0.028$),  3C 205 ($z=1.534$), and 4C 41.17 ($z=3.8$).  The lower limit in this range arises from the suggestion that the jet may dither in its direction as it passes through hot spot backflow material close to the jet termination point, creating a $``$dentist drill" effect on the inside of a cavity 700 pc in diameter.
\end{abstract}

\keywords{Galaxies: active, Galaxies: individual (Pictor A), Galaxies: jets, Radio continuum: galaxies}

\section{Introduction}

The hot spots of powerful radio galaxies, where relativistic jets originating at the active galactic nucleus (AGN) terminate in an interaction with the intergalactic medium (IGM), are known to radiate strongly over the radio to X-ray wavelength range (\cite{geo03} list some of the best studied objects).  At radio wavelengths the emission mechanism is widely agreed to be due to the synchrotron process.  The details of the emission mechanisms responsible for the production of the X-rays in the jets and hot spots of powerful radio galaxies are a matter for more debate \citep{har06,geo03,tav00}, with the relative contributions of synchrotron emission and the various flavours of inverse Compton emission (synchrotron self-Compton and external Compton) often difficult to determine from the models.  Complicating factors include the orientation of the jet relative to the observer and the degree of Doppler boosting from the different regions of the jet and hot spot, caused by possibly relativistic flows in these regions.

Although our knowledge of the high energy emission from radio galaxy hot spots has greatly improved since the launch of \emph{Chandra} \citep{sch00}, and knowledge of the synchrotron emission at low photon energies has been very good for many years from instruments such as the VLA and large optical/infrared telescopes, little is still understood of the structure of the hot spots at the highest possible spatial resolutions.  As the hot spot emission is widely considered to originate in strong shocks produced when the jet interacts with the IGM, high resolution observations have the potential to determine the structure of these shocked regions.  Very long baseline interferometry (VLBI) is the highest resolution direct imaging technique in astronomy and can produce milliarcsecond resolution images of the radio emission from radio galaxy hot spots.  Important parameters in models for the X-ray emission from the hot spots can be estimated from these observations, such as the volume and energy density of the shocked regions.  

Despite this, the literature contains few reports of VLBI observations of radio galaxy hot spots.  \cite{kap79} presented single-baseline VLBI observations of hot spots in 35 high luminosity, distant ($z>0.5$) type II quasars or radio galaxies, showing that structures more compact than 150 mas can exist.  \cite{lon98}, \cite{lon89}, and \cite{lon84} report on VLBI observations of a hot spot in 3C 205 at a redshift of 1.534, finding a compact and complex structure $\sim$300 mas in extent (corresponding to $\sim$2.5 kpc\footnote{In this paper we adopt a cosmology with $H_{0}=71$ km/s/Mpc, $\Omega_{m}=0.27$, and $\Omega_{\Lambda}=0.73$ \citep{spe03}.  We calculate all parameters for Pictor A using this cosmology and recompute previously published results in the literature also according to this cosmology, for comparison.}), observing with an angular resolution of 6.9 $\times$ 8.8 mas (corresponding to a spatial resolution of approximately 60 pc).  The comprehensive investigation of 3C 205 led \cite{lon98} to the conclusion that the complex structure seen in the hot spot is due to a continuous jet flow around a bend caused by interaction with a dense medium.  

\cite{gur97} report on VLBI observations of the hot spot in 4C 41.17 ($z = 3.8$) that show compact structure with not more than 15 mas angular extent ($\sim$110 pc at this redshift), accounting for $\sim$30\% of the integrated flux density of the hot spot as imaged with the VLA.  \cite{gur97} conclude that the structure detected with VLBI is the location of an interaction with a massive clump in the interstellar medium, causing a deflection in the jet direction.

Most recently, \cite{you05} report on VLBI observations of the southern lobe hot spot of the nearby radio galaxy PKS 2153-69 ($z=0.028$), showing that, at 90 $\times$ 150 mas resolution, the hot spot is marginally resolved into three circular Gaussian components of 100 - 220 mas FWHM and 10 - 65 mJy, within a 400 mas diameter area.  The spatial resolution of the PKS 2153-69 observations were comparable to the 3C 205 observations of \cite{lon98}, approximately 50 pc.

Pictor A is the closest powerful Fanaroff-Riley type II (FR-II) radio source, at a redshift of 0.03498(5) \citep{era04}.  By virtue of its proximity, Pictor A is bright over a wide wavelength range, from radio to X-ray, and offers good spatial resolution return for high angular resolution (1 milliarcsecond $=$ 0.7 pc).  Previously, a number of authors have taken advantage of Pictor A (and its host galaxy) as a target for detailed observations on a variety of spatial scales, at radio \citep{per97,tin00,sim99}, optical \citep{sim99}, and X-ray \citep{wil01} wavelengths.  \cite{wil01} undertook an exploration of the emission mechanisms possible for the hot spot X-rays, arriving at a number of plausible explanations, involving several distinct physical interpretations.  Their analysis highlights the general uncertainties in models for X-ray emission from hot spots noted above.  Better knowledge of the high resolution structure of the Pictor A hot spots may help our understanding of the X-ray emission mechanisms at play, which may be useful for our general understanding of radio galaxies.

In this paper, we present VLBA observations of the north-west Pictor A hot spot that reveal complex and compact structure on an angular scale of 23 milliarcseconds, corresponding to a spatial scale of 16 pc, resolving the termination shock of the jet at the IGM.  We make a comparison of the VLBA data to the VLA data of \cite{per97}, the ATCA data of \cite{len08}, the \emph{Chandra} data of \cite{wil01}, HST data of \cite{mei97} and new Spitzer infrared data, discussing possible physical explanations for the high resolution structure of the hot spot and implications for synchrotron and inverse Compton modelling of radio galaxy hot spots.  The VLBA image of the Pictor A hot spot is the highest spatial resolution image of a radio galaxy hot spot to date.

\section{Observations and results}

\subsection{VLBA observations}

Pictor A was observed at 18 cm and 13 cm using the NRAO Very Long Baseline Array (VLBA) on 2005 November 20 and 21, respectively. Due to the low declination of the target, $-45\arcdeg$, only the southern antennas of the array at Los Alamos, Fort Davis, Pie Town, Kitt Peak, and Owens Valley were used. The observation spanned approximately 5 hours with 10 minute scans of the fringe finder PKS 0537$-$441 made on an hourly basis. Dual circular polarisation data were recorded across four 4 MHz intermediate frequencies (IFs) centred on 1653.48, 1661.48, 1669.48 and 1677.48 MHz for the 18 cm observation and 2257.49, 2265.49, 2273.49 and 2281.49 MHz for the 13 cm observation. The data were correlated at the NRAO VLBA correlator (Socorro, NM, USA) with the AGN core as the phase centre. To reduce the effects of bandwidth smearing and time averaging smearing at the north-west and south-east hot spots, in order to make a wide field of view available for imaging, the correlator generated data with 128 spectral points per baseline/IF and an integration time of 0.26 s. The one sigma theoretical thermal noise of the 18 cm and 13 cm data sets is 170 $\mathrm{\mu}$Jy beam$^{-1}$ and 190 $\mathrm{\mu}$Jy beam$^{-1}$, respectively.

Initial calibration of the data was performed using standard VLBA data calibration techniques in AIPS\footnote{The Astronomical Image Processing System (AIPS) was developed and is maintained by the National Radio Astronomy Observatory, which is operated by Associated Universities, Inc., under co-operative agreement with the National Science Foundation}. 

The calibration was refined by using the nucleus of Pictor A as an in-beam calibrator. To account for structure in the nucleus (which we use as a phase reference source) a new DIFMAP \citep{she94} task, \emph{cordump}\footnote{The \emph{cordump} patch is available for DIFMAP at \url{http://astronomy.swin.edu.au/$\sim$elenc/DifmapPatches/}} \citep{len06}, was developed to enable the transfer of all phase and amplitude corrections made in DIFMAP during the imaging process to an AIPS compatible solution table. The \emph{cordump} task greatly simplified the calibration of the data set. 

First, the data were averaged in frequency and exported to DIFMAP where several iterations of modelling and self-calibration of both phases and amplitudes were performed. \emph{cordump} was then used to transfer the resulting phase and amplitude corrections back to the unaveraged AIPS data set. The bandpass for the data was calibrated against observations of PKS $0537-441$. After application of these corrections, the DIFMAP model of the nucleus was subtracted from the unaveraged (u-v) data set. The 18 cm and 13 cm images of the nucleus had a measured RMS noise of 240 and 500 $\mathrm{\mu}$Jy beam$^{-1}$, respectively. The higher than theoretical noise in the 13 cm data is attributed to substantial levels of RFI observed at Kitt Peak and Los Alamos during the observation.

At 18 cm, the nuclear core total flux density is 734 mJy and is composed of a 526 mJy ($27\times11$ mas at a position angle of $-8\arcdeg$) component and a 209 mJy ($62\times38$ mas at a position angle of $-61\arcdeg$) extension to the north-west.  At 13 cm, the nuclear core total flux density is 885 mJy and is composed of a 624 mJy ($18\times5$ mas at a position angle of $-36\arcdeg$) component and a 260 mJy ($110\times62$ mas at a position angle of $10\arcdeg$) extension to the north-west.  Figure 1 shows the 18 and 13 cm images of the AGN in Pictor A.  The resolution of these images is significantly worse than the 3 cm images in \cite{tin00}, so a detailed comparison with those previous results is not possible.  The unaveraged, calibrated, core-subtracted datasets were then used to image the regions around the north-west and south-east hot spots in Pictor A.

In the first phase of the hot spot imaging process, the AIPS task IMAGR was used to make naturally weighted dirty images and beams of regions covering both the south-east and north-west hot spots. A $3\times3$ grid of $\sim12\arcsec$ square dirty maps were imaged simultaneously using the multi-field option within IMAGR.  The DO3D option, in combination with the gridded imaging, was used to reduce non-coplanar array distortion. Each grid was centred about the brightest components of each hot spot, as determined from lower resolution 6 cm ATCA (Australia Telescope Compact Array) images. Since each dirty map contains $\sim10^5$ synthesized-beam areas, a conservative $6\sigma$ detection threshold was imposed to avoid spurious detections. Furthermore, only the inner 75\% of each dirty map was searched for candidate detections to avoid erroneous detections as a result of map edge effects.

The second phase of the imaging process involved creating a (u,v) shifted data set for each of the positive detections, using the AIPS task UVFIX. The shifted data sets were averaged in frequency, effectively reducing the field of view of each of the targeted sources to approximately $10\arcsec$, and then exported to DIFMAP. In DIFMAP, the visibilities were averaged over 10 second intervals to reduce the size of the data set and to speed up the imaging process. Each target was imaged in DIFMAP, with natural weighting applied, using several iterations of CLEAN.

Earlier observations of Pictor A with the VLA (Perley et al. 1997) determined the position of the AGN core as $\alpha=05\rah18\ram23\fs590$ and $\delta=-45\arcdeg49\arcmin41\farcs40$ (B1950 epoch 1979.9) or $\alpha=05\rah19\ram49\fs706$ $\delta=-45\arcdeg46\arcmin43\farcs44$ (J2000). To provide a direct comparison with the VLA data we have re-referenced our data so that our nuclear core coincides with the VLA position.


No sources were detected in the hot spot regions at 13 cm. Only the north-west hot spot was detected at 18 cm.  Given the peak flux density we observe at 18 cm, the expected hot spot spectral index leading to weaker emission at 13 cm, the smaller beam size at 13 cm, and the less sensitive VLBA image at 13 cm, we calculate that the expected 13 cm peak flux density would lie below our detection limit.

Figure 2 shows the VLBA image of the north-west Pictor A hot spot at 18 cm.  The image consists of a number of components.   Figure 3 shows the final VLBA image of the hot spot, overlaid on a VLA image at 15 GHz (Figure 18 from \cite{per97}), showing the hot spot within the context of the overall structure of the Pictor A radio source, using VLA data from \cite{per97} and ATCA data from \cite{len08}.

In our 18 cm observation, sources approximately 4$\arcmin$ from the phase centre, the approximate distance to the south-east and north-west hot spots, exhibit a 6.4\% loss in peak flux density as a result of the cumulative effect of bandwidth smearing (1.6\%), time averaging smearing (0.8\%) and primary beam effects (4.1\%).  Our measured flux densities were corrected for primary beam effects (4.1\% adjustment to flux densities), but not corrected for smearing effects, due to the intrinsic extended nature of the components seen in the hot spots.  The combined 1.6\% and 0.8\% (2.4\%) effect of smearing has been included in the error estimates for the hot spot flux densities.  The errors on the peak and integrated flux densities are therefore the quadrature combination of the smearing effects and the absolute error in the calibration of the flux density scale, $\sim$10\%.

The parameters of the components (as annotated in Figure 2) are listed in Table 1, adjusted for the primary beam effects as outlined above.

The inherent positional uncertainty in both the VLA and VLBA images is approximately 50 mas.  The nucleus in the VLBA image is manually aligned with the VLA coordinates of the nucleus, which have approximately 50 mas uncertainty \citep{per97}.  Although the positional accuracy of the VLA image of the hot spot at 15 GHz is not stated in \cite{per97}, since the hot spot image was referenced to the same calibrator and self-calibrated, presumably the positional accuracy of VLA image of the hot spot is also of order 50 mas.  The alignment between the VLA and VLBA images is therefore probably good at the 100 mas level. 

The positional uncertainties, combined with the difference in angular resolution and observing frequency between the VLA data and the VLBA data, make a detailed registration of the two datasets difficult.  The VLBI components trace the bright part of the VLA image but no one component marks the peak of the brightness distribution in the VLA image.  The brightest of the VLBA components lies furthest from the core, beyond the peak in the VLA image, in the jet direction.  It appears that the structure seen at lower resolution with the VLA may therefore be a combination of the compact components seen with the VLBA, combined with diffuse emission to which the VLBA observations are not sensitive, since the sum of the flux densities of the components seen with the VLBA are a small fraction of the VLA flux density in the hot spot at the same frequency.  The integrated flux density in the VLBA image is approximately 120 mJy, whereas the integrated flux density at 1.7 GHz within the 1\arcsec~ region encompassing the emission seen with the VLBA is close to 8 Jy (based on an interpolation of the radio data in \cite{per97}).  The unseen diffuse emission must trace the compact components however, as the compact components follow the higher contours of the VLA image.  This implies that the emission seen with the VLBA represents the high brightness temperature $``$cores" of the hot spot sub-components.

Figure 4 shows a comparison between the VLBA data and the optical data of \cite{tho95}.    The raw optical image was retrieved from the HST archive and a Gaussian smoothing function (with a kernel radius of 10) was applied to the image using DS9 \citep{joy03} before comparing to the VLBA image.  The peak of the optical image (uncalibrated intensity) was aligned with the peak of the 15 GHz VLA image of \cite{per97}, in order to register against the VLBA data.  Again, significant positional uncertainties exist between the HST image and the VLBA image, at least of order 100 mas, due to the alignment of the peaks in the HST image and the VLA 15 GHz image.  Likely this HST-VLBA registration is significantly more uncertain than at the 100 mas level because of the frequency difference between the HST and VLA 15 GHz images.

\subsection{SPITZER observations}

The north-west hot-spot in Pictor A was observed by Spitzer with IRAC (Infrared Array Camera) and MIPS (Multiband Imaging Photometer for SIRTF).  The IRAC observations were made on 2004 November 26 with a total time of 48 s. The MIPS (24 $\mu$m only) observations were made on 2004 September 21 with an observation time of 100 s. The images were produced using the data analysis tool MOPEX \citep{mak06}.  The IRAC and MIPS fluxes were obtained from circular regions with radius of 3.6 arcsec and 6 arcsec, respectively, centered on the radio peak of the hot-spot. These values were corrected applying the appropriate aperture correction for the adopted extraction region.  Table 2 lists the Spitzer data.

\section{Discussion}

We have detected only the north-west hot spot in Pictor A at radio wavelengths with the VLBA, thus this hot spot contains higher brightness temperature components than the south-east hot spot.  This is consistent with the ideas put forward by \cite{geo03}, that radio galaxies with somewhat aligned jets (they count Pictor A among these objects) have brighter hot spots on the side of the approaching jet, due to Doppler boosting of the emission (assuming that the hot spots are identical).  The north-west hot spot brightness temperature may therefore be enhanced by Doppler boosting compared to the south-east hot spot, making it easier to detect with VLBI.  This is consistent with the jet sidedness as seen in VLBI images of the AGN core \citep{tin00} and the \emph{Chandra} X-ray data of \cite{wil01}.  Alternatively or additionally, the south-east hot spot of Pictor A has a significantly different morphology to the north-west hot spot, a double hot spot morphology similar to structures observed in a number of other FR-II radio galaxies e.g. \cite{lon89}.  In the south-east hot spot, the jet may release energy in multiple weaker interactions, rather than in a single strong interaction as appears in the north-west hot spot.  The nature of the south-east hot spot is briefly discussed below.

\subsection{Implications for X-ray emission models}

\subsubsection{Previous models - balanced synchrotron and SSC emission}

In their exploration of various models for the X-ray emission from the north-west hot spot in Pictor A, \cite{wil01} consistently assume a radius for the emission region of 250 pc = 0.4\arcsec, based on the VLA data.  As is usual in such models, the assumption is made of a homogenous, spherical emission region, which does not appear to be a bad assumption based on the VLA data, alone.  The VLBI data, however, show the danger in such an assumption, revealing significant structure on scales more than an order of magnitude smaller than the assumed size.  Further modelling is therefore required to assess the X-ray emission from a combination of compact and diffuse structures, to determine deviations from the models presented in \cite{wil01}, and we explore these models in detail below.  It is worth noting that, although a relatively small percentage of the integrated flux density of the hot spot is contained in the VLBI components, significant structure is likely to be present on spatial scales that fall between the VLA and VLBA resolutions, to which neither array is sensitive.  It is therefore possible that the majority of the flux density of the hot spot may be contained in sub-components that are a factor of 2 or 3 smaller than assumed by \cite{wil01}.  Given the resolved nature of the structures we detect with the VLBA, it is unlikely that a significant portion of the detected flux density contained in these structures exists in unresolved components.  We estimate conservatively that less than 10\% of the detected flux density could be contained in unresolved components.  Observations on longer and more sensitive VLBI baselines would be required to quantify this statement further.

\cite{wil01} note that the X-rays could be due to synchrotron radiation and find that their synchrotron model is consistent with particle acceleration in strong shocks, the electrons requiring reaccelaration in shocks on pc scales.  The components seen in our VLBI images are on the right spatial scales to be consistent with this picture.

Therefore, both the possibilities of a smaller emission region for the X-rays, and the generation of synchrotron X-rays could be explored further in models for the north-west Pictor A hot spot.  Interestingly, \cite{wil01} find that a composite synchrotron plus synchrotron self-Compton (SSC) model can match the \emph{Chandra} observations, but requires similar contributions from both processes in the \emph{Chandra} band.

However, we believe that these models are not strongly predictive, as  \cite{wil01} model the synchrotron peak as being due to a broken powerlaw electron distribution, giving the synchrotron tail required in the \emph{Chandra} band, with the SSC peak derived from an electron energy distribution that cuts off sharply at a maximum energy.  The addition of the synchrotron and SSC peaks is therefore based on two different electron energy distribution models that are qualitatively similar, but strictly inconsistent.

\subsubsection{New models - synchrotron dominated emission}

With this in mind, and the new information available from the VLBA observations, we have endeavored to find a more natural explanation for the radio to X-ray spectrum for the Pictor A hot spot.  The pc-scale structures seen in the hot spot are compact regions with energy density 2--3 times larger than that of the (average) hot spot, thus they may trace the region where electrons are accelerated very recently via shocks and turbulence.  These should be transient regions that are expected to expand reaching 
pressure equilibrium with the surrounding plasma. Their maximum dynamical time--scale can be estimated by the Alfvenic crossing time or by the time required (with an advance speed $\approx$ 0.05--0.1 $c$) to cross these regions, and this comes out to be between $\sim$5 years and a few decades.  This is much smaller than the radiative time scale of the overall diffuse hot spot, that is about 100-700 years, assuming a break frequency of the synchrotron spectrum, $\nu_b \approx 10^{14}$ Hz (resulting from the frequency of the peak of the synchrotron emission originating in the diffuse region surrounding the pc-scale hot spot structures: Figure 5) , and a magnetic field in the range $100-400~\mu$G (this range being consistent with the expected equipartition magnetic field strength).  Thus the frequency of the radiative break in the synchrotron spectrum of the dynamically young regions discovered by our VLBA observations, that scales with $\nu_b \propto B^{-3} \tau^{-2}$, where $\nu_b$ is the break frequency, $B$ is the magnetic field strength and $\tau$ is the timescale, is expected to be $\approx$100--10000 times higher than that of the overall hot spot.  Therefore the synchrotron emission contribution from the pc--scale components would extend into the \emph{Chandra} band.

Based on these considerations we adopt a model of the hot spot region assuming two emitting components : the diffuse contribution to the hot spot, that dominates in the radio and optical band, and the pc--scale components embedded within the diffuse emission, that dominates the X-ray emission.  As an approximation, each component (diffuse and each pc--scale component) is described by a homogenous sphere with constant magnetic field and fixed properties of the relativistic electrons.  We model the spectral energy distributions of the emitting electrons in both the diffuse and the pc--scale components by means of the formalism described in \cite{bru02} that accounts for the acceleration of electrons at shocks and for the effect due to synchrotron and inverse Compton losses on the shape of the spectrum.

First we fit the synchrotron emission of the diffuse emission using the radio, IR and optical data points (using same data as used by \cite{wil01}, plus our new Spitzer data) and derive the relevant parameters of the synchrotron spectrum (injection spectrum $\alpha$, break frequency $\nu_b$, cut-off $\nu_c$) and the slope of the spectrum of the electrons as injected at the shock ($\delta = 2 \alpha +1$).  For a given value of the magnetic field strength this allows us to fix the spectrum of the emitting electrons (normalization, break and cut-off energy), and to calculate synchrotron self-Compton (SSC) emission from the hot spot region, following \cite{bru02}.

The synchrotron spectrum of the pc--scale components is normalised to the fluxes at 18 cm derived from our VLBA observations, assuming the same $\alpha$ measured for the emission from the diffuse region and a synchrotron break at higher frequencies.  For a given value of the magnetic field strength in the region of the pc--scale components the spectrum of the electrons is fixed and the inverse Compton emission from these electrons is calculated by taking into account both SSC and the scattering of the external radio photons from the diffuse region in which the pc--scale components are embedded.

In Figure 5 we show the radio to hard X-ray emission expected from our modelling of the hot spot region of Pictor A.  Synchrotron emission is responsible for the observed properties of the hot spot, while the inverse Compton emission is expected to give appreciable contribution only at very high energies.  In Figure 5 the SSC is calculated assuming a reference value B=350 $\mu$G in both the diffuse and pc--scale components \footnote{this value is compatible (within a factor of 2) with the equipartition field in both these regions calculated with equipartition formulae given in \cite{bru97} with $\gamma_{min}=100$ and F=1.} with the two components giving a similar contribution to the total SSC spectrum.

As noted in \cite{wil01}, an assumed magnetic field an order of magnitude lower than the equipartition field is required to reproduce the observed X-ray luminosity, but fails to reproduce the X-ray spectral slope.

Thus, our new observations have given a basis for a synchrotron X-ray component which, when self-consistently calculated along with a SSC X-ray component, shows that the \emph{Chandra} band is dominated by synchrotron emission from the pc-scale components, with a relatively small SSC contribution.  The implication of this is that at energies higher than observed with \emph{Chandra}, the new model predictions stand in stark contrast to those of \cite{wil01}, providing a clear test of the two models in the future.

The new synchrotron-dominated model provides a much more natural explanation for the Pictor A hot spot radio to X-ray emission, taking into account the new VLBA observations of significant pc-scale internal structure.  The model calculation is also fully self-consistent, compared to the previous models, so we believe that there is good reason to suppose that our analysis reflects closely on the true physical situation for the hot spot.  Further higher energy observations with good angular resolution and sensitivity are required to ultimately discriminate between the very different predictions of our model and those of \cite{wil01}.

\subsubsection{Other considerations}

\cite{geo03} raise the possibility that the bulk flow of the material in radio galaxy hot spots, in the lab frame, may be at least mildly relativistic and therefore allow inverse Compton scattering of the cosmic microwave background as a viable model for the X-ray emission.  The post-shock flow may be relativistic \citep{geo03} and the hot spot advance into the IGM may be mildly relativistic (\cite{wil01}; \cite{geo03}).  Of particular interest, therefore, is any measurement of the hot spot advance speed in Pictor A, which could strengthen or weaken this argument.  The mismatch in resolution between the VLA and VLBA images in Figure 3, and the sum of the positional uncertainties in both images, do not allow any meaningful constraint on the hot spot advance speed.  Further VLBI observations are required at much later epochs, to determine any hot spot advance.  \cite{ars00} find an average value of $v_{adv} \sim 0.1c$ for powerful radio galaxies.  Given this value at the distance of Pictor A, $\sim$0.05 mas/yr could be expected for the hot spot advance.  Even over a 20 year period, only of order a milliarcsecond of difference in angular position could be expected.  At the resolution and sensitivity of current VLBI observations, such a detection of motion would be very difficult.  Any changes in the $\sim$15 years between the VLA and VLBA observations are therefore insignificant compared to the resolution of either observation.

\subsection{Terminal shock morphology and comparison to 3C 205, PKS 2153$-$69, and 4C 41.17}

How does the distribution of presumably shocked jet material seen in the VLBA image of the Pictor A hot spot relate to the termination of the jet at the IGM?  The $``$dentist drill" scenario explored by \cite{sch82} and \cite{cox91} means that the average jet power released in the jet termination is spread over a larger area than the crossection area of the jet.  Although there would only be a single termination shock at any one time, the locations of previous termination shocks would also be sites of radio emission, until their electron populations lost enough energy to be undetectable at radio wavelengths.  However, the $``$dentist drill" model is typically used to explain hot spot morphologies on much larger scales ($\sim$kpc) than probed with our VLBA data.  Additionally, the $``$dentist drill" model, as simulated by \cite{cox91} was caused by a substantial precession of the jet (5 - 10$^{\circ}$ half-angle variations of the jet direction), manually introduced into the simulation.  No such evidence for precession of the north-west jet in Pictor A can be seen, since the VLBI, VLA, and \emph{Chandra} data indicate a very straight jet that originates at the nucleus and terminates at the hot spot.

A small scale analog of the $``$dentist drill" effect may be possible.  As the straight jet approaches the termination point, traversing the relatively high density backflow from the termination point, the direction of the jet may be altered, causing the termination point to vary over a working surface larger than the cross-section of the jet.  Assuming that this is the case, and each of the components seen in Figure 2 are current or previous interaction sites, it implies that the jet width is of order 100 mas $=$ 70 pc, the observed size of the components in the VLBI image.  Optical observations of the hot spots in 3C 445 showing clumps of emission proposed as ``local accelerators" could also be interpreted under a small-scale dentist drill scenario \citep{pri02}.  A 70 pc jet width at the hot spot implies a very small jet opening angle, $<$0.2$^{\circ}$.

Alternatively, the shocks may simply be the highest brightness temperature regions of an extended termination shock front, which is spread over a larger area than indicated by each of the compact components.  In this case, the spread of components may be a better indication of the crossection of the jet at the hot spot, 1000 mas $=$ 700 pc.  In this case, the distribution of compact components may indicate the structure of the IGM that the jet is interacting most strongly with.

Comparisons of the VLBA image to both the VLA image and the HST image of the hot spot shows that the compact structures are distributed in a symmetric fashion around the jet direction.  The direction of the nucleus at the hot spot is indicated in both Figures 3 and 4 and the two alternatives discussed above cannot be distinguished on the basis of the distribution of the compact emission.

We note, in comparison, that the overall size of the hot spot detected with VLBI in 3C 205, by \cite{lon98}, is 1400 pc, although the morphology is quite different to that seen in Pictor A.  The brightest, most compact part of the 3C 205 hot spot (Component 3 in the \cite{lon98} nomenclature) appears to be of order 250 pc in size and is much more luminous than the hot spot in Pictor A.  The conclusion that \cite{lon98} apply to the 3C 205 data, of a continuously bending jet in reaction to the jet interaction with a dense IGM, does not seem to immediately apply to the Pictor A data.  There is no clear connection between the components in Figure 2 that would indicate a flow of material from one component to another.  In this sense, the small-scale $``$dentist drill" model may be more plausible.

In 4C 41.17, the observed compact structure in the hot spot at the suggested site of a jet interaction has a size of $\sim$110 pc, comparable to that seen for the individual components in the VLBA image of the Pictor A hot spot.

The only other well observed example of a radio galaxy hot spot at VLBI resolution is PKS 2153$-$69 \citep{you05}.  In PKS 2153$-$69, the overall hot spot size is approximately 200 pc, containing structures as small as 50 pc in extent, comparable to the size of the structures seen in 3C 205 and Pictor A.   Interestingly, in PKS 2153$-$69, the hot spot structure on scales of $\sim$5 kpc shows evidence for a similar morphology as seen in 3C 205, a curved structure with a strong surface brightness gradient.  In PKS 2153$-$69, the overall radio morphology, along with X-ray and optical data are interpreted to support a model of jet precession, where the curved radio structure in the hot spot traces the varying position of the jet interaction region with time, rather than an {\it in situ} bend of the flow due to the jet interaction.  This is the type of radio galaxy structure that the classical $``$dentist drill" model of \cite{sch82} seeks to explain.  Again, the current data for Pictor A do not seem to easily fit such an interpretation.

\subsection{Comparison to jet simulations}

\cite{sax02} present two-dimensional, axisymetric, non-relativistic, hydrodynamic simulations of the north-west Pictor A hot spot, in an effort to reproduce the distinctive hot spot structure on arcsecond scales (in particular the filament behind the hot spot that appears perpendicular to the jet direction).  Although these simulations are simplistic relative to the physical situation in Pictor A, they show that the hot spot advance proceeds in an episodic and ephemeral manner, driven by the formation of a channel or cavity evacuated by the jet, the temporary frustration of the jet as the channel periodically closes, and a rapid advance of the hot spot as it breaks through the temporary obstacle, traversing the channel rapidly and reforming the working surface at the IGM.  

The simulations show that a rich variety of complex structures are possible at the terminal shock of the jet, although the simulation resolution does not reach the equivalent of the mas scales of Figure 2.  The estimated timescales for these predicted temporal variations in the hot spot position ($10^{4} - 10^{5}$ years, depending on the simulation) are much longer than the period between the VLA and VLBA images for Pictor A and the observations would not be expected to detect these changes in the hot spot position with time.

\subsection{The south-east hot spot in Pictor A}

Although the south-east hot spot in Pictor A was not detected with our VLBA observations, it provides an interesting contrast to the north-west hot spot.  The south-east hot spot appears to be a double hot spot, as noted in a number of other radio galaxies.

The classical $``$dentist drill" model, as simulated by \cite{cox91} invokes jet precession to provide position angle changes of the jet, allowing the termination point of the jet to vary across an area larger than the width of the jet.  In Pictor A, there is no evidence for jet precession in the north-west jet, and therefore not likely to be any precession of the south-east jet which feeds the south-east lobe.  We therefore find that the $``$dentist drill" model is not likely to explain the morphology of the south-east lobe.  

The jet bending hypothesis put forward by \cite{lon98} to explain the double hot spot in 3C 205 is supported by the evidence that the jet from the nucleus in this object connects with a curved, bright hot spot that subsequently appears to feed a weaker hot spot further from the nucleus.

In Pictor A, such a situation is not apparent.  Figure 6 shows an overlay of radio contours from an ATCA image \citep{len08} with co-added images from the \emph{Chandra} archives (also published in \cite{har05}).  The yellow line in the figure indicates the direction of the X-ray and nuclear VLBI jets, which shows they are highly aligned with the north-west hot spot that we have detected with VLBI.  This jet direction, extrapolated to the counterjet side, is misaligned with the south-east double hot spot, by approximately 20$^{\circ}$.  There is no point along this extrapolated jet direction that provides obvious evidence of an interaction that would bend the jet and deflect it to the location of the observed double hot spots.  A relatively strong X-ray source appears aligned with the extrapolated jet direction, immediately before the double hot spots, however, there is no radio counterpart to the X-ray component (c.f. Figure 3).  The observations therefore do not obviously support the jet bending hypothesis \cite{lon98} used for 3C 205.

The nature of the south-east hot spots in Pictor A, and the apparent misalignment with the jet, relative to the north-west jet and hot spot is therefore an interesting aspect of this galaxy, requiring further observation and interpretation, in particular very deep radio and X-ray imaging in order to trace the jet on its path to the south-east hot spots.  The \emph{Chandra} co-added image shown in Figure 6 is more sensitive on the north-west side of the galaxy than on the south-east, as this is where strong emission was known to exist and the observations concentrated.  Further \emph{Chandra} observations of the region of the south-east jet, hot spots, and lobes would be very interesting.

\section{Conclusions}

Of the four radio galaxy hot spots that have been imaged in detail using VLBI techniques, all have similar gross properties in that they possess structure on scales between 20 and 100 pc, despite being at redshifts of 0.028, 0.035, 1.534, and 3.8.  The details of the structures observed in the different hot spots are rather different, however, and different interpretations have been applied to these structures, from the bent jet model of \cite{lon98} to the precessing jet model of \cite{you05}, and the jet interaction conclusion of \cite{gur97}.  A small-scale analog of the $``$dentist drill" model may be a viable scenario for the north-west hot spot Pictor A, but other scenarios are also possible.  

The detection of parsec-scale structures in radio galaxy hot spots has implications for the models used to explain the X-ray emission from the hot spots.  In particular we find that a natural explanation for the radio to X-ray spectrum is that the pc-scale components in the hot spot represent recently accelerated regions of electrons, with higher break frequencies than the remainder of the hot spot electrons, on larger spatial scales.  A consequence of this is that the X-rays in the \emph{Chandra} band are dominated by synchrotron X-rays, with a relatively weak SSC contribution.  This is in stark contrast to the previous modelling of \cite{wil01}, who predict a balanced SSC and synchrotron emission in the \emph{Chandra} band and strong SSC emission above the \emph{Chandra} band.  Future high energy observations may readily distinguish between these models.

The south-east hot spot has a substantially different morphology to the north-west hot spot, which may have to do with the apparent misalignment between the jet and the south-east hot spot location.  Further observations, particularly at X-ray wavelengths would help to determine the nature of the south-east hot spots.

Obviously, further systematic VLBI observations of hot spots in other nearby powerful radio galaxies would be highly useful, to determine if similar high resolution structures exist in those sources.  FR-II radio galaxies at low redshift, where high spatial resolution can potentially be obtained, are rare.  We have therefore embarked on a program to observe the hot spots in a number of powerful radio galaxies with $z<0.1$ in the Southern Hemisphere.  

A further potentially interesting high angular resolution observation of the Pictor A hot spot has been proposed by \cite{wil01}.  Their modelling hints at the presence of a low energy electron population in the hot spot that is not visible at cm wavelengths.  To detect such a population, low frequency observations with high resolution would be required.  At some point in the future, the Murchison Widefield Array (MWA), being built by an international consortium at the candidate Square Kilometere Array (SKA) site in Western Australia, may be able to make such an observation.  Also possible would be to observe the Pictor A hot spots with the 90 cm system on the VLBA, to obtain very high resolution at long wavelengths.  Although too far north to observe Pictor A, the eLOFAR array will have good angular resolution at very low radio frequencies and will be able to observe the hot spots of many Northern Hemisphere radio galaxies to search for low energy electron populations, such as suggested by \cite{wil01} for Pictor A.

\acknowledgments
We thank Rick Perley for providing VLA images for the Pictor A hot spot used in Figure 3.  EL acknowledges a Swinburne University of Technology Chancellor's Research Scholarship and a CSIRO top-up scholarship.  The National Radio Astronomy Observatory is a facility of the National Science Foundation operated under cooperative agreement by Associated Universities, Inc.  This research has made use of NASA's Astrophysics Data System.  This research has made use of the NASA/IPAC Extragalactic Database (NED) which is operated by the Jet Propulsion Laboratory, California Institute of Technology, under contract with the National Aeronautics and Space Administration.  Some of the data presented in this paper were obtained from the Multimission Archive at the Space Telescope Science Institute (MAST). STScI is operated by the Association of Universities for Research in Astronomy, Inc., under NASA contract NAS5-26555. Support for MAST for non-HST data is provided by the NASA Office of Space Science via grant NAG5-7584 and by other grants and contracts.

{\it Facilities:} \facility{VLA}, \facility{VLBA}, \facility{ATCA}.

\clearpage

\begin{figure}
\plottwo{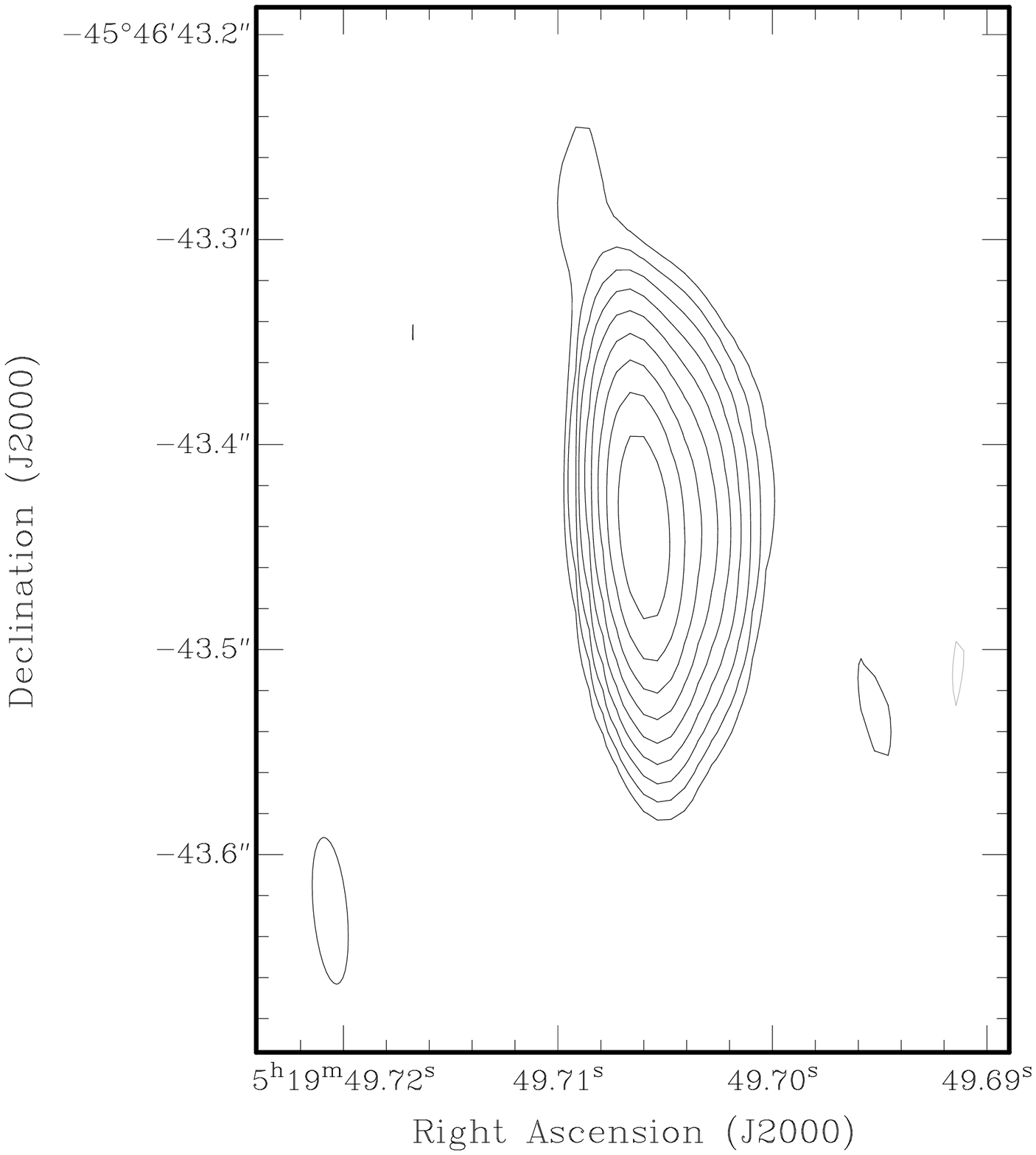}{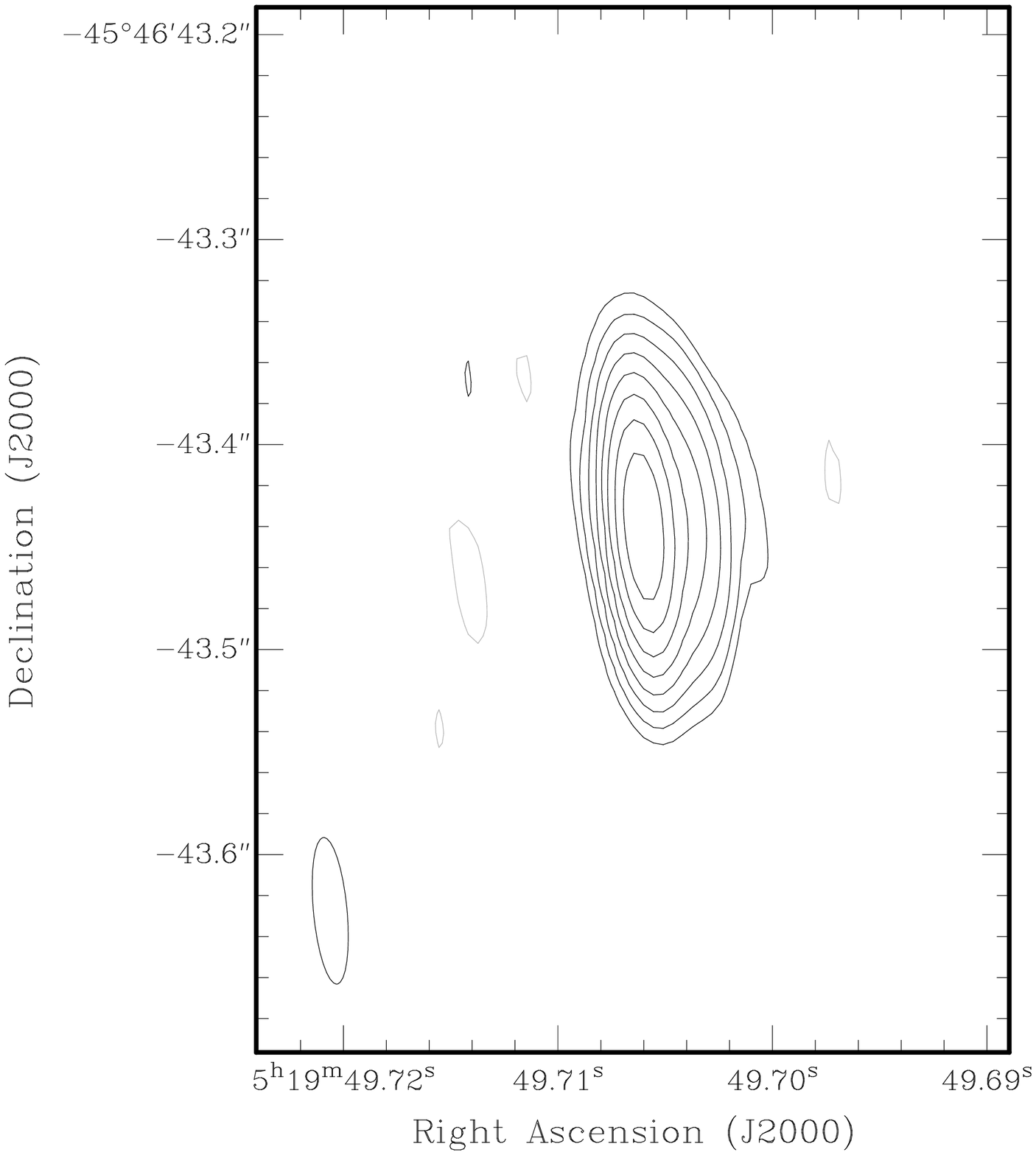}
\caption{VLBA images of the Pictor A nucleus at 18 cm (left) and 13 cm (right).  At 18 cm the image has an RMS noise level of approximately 240 $\mathrm{\mu}$Jy beam$^{-1}$. The peak flux density is 592 mJy beam$^{-1}$ and contours are set at -0.2\%, 0.2\%, 0.4\%, 0.8\%, 1.6\%, 3.2\%, 6.4\%, 12.8\%, 25.6\% and 51.2\% of the peak. The beam size is approximately $94\times22$ at a position angle of $5\arcdeg$.  At 13 cm the image has an RMS noise level of approximately 500 $\mathrm{\mu}$Jy beam$^{-1}$. The peak flux density is 654 mJy/beam and contours are set at -0.4, 0.4\%, 0.8\%, 1.6\%, 3.2\%, 6.4\%, 12.8\%, 25.6\% and 51.2\% of the peak. The beam size is approximately $72\times16$ at a position angle of $5\arcdeg$.\label{fig1}}
\end{figure}

\clearpage

\begin{figure}
\plotone{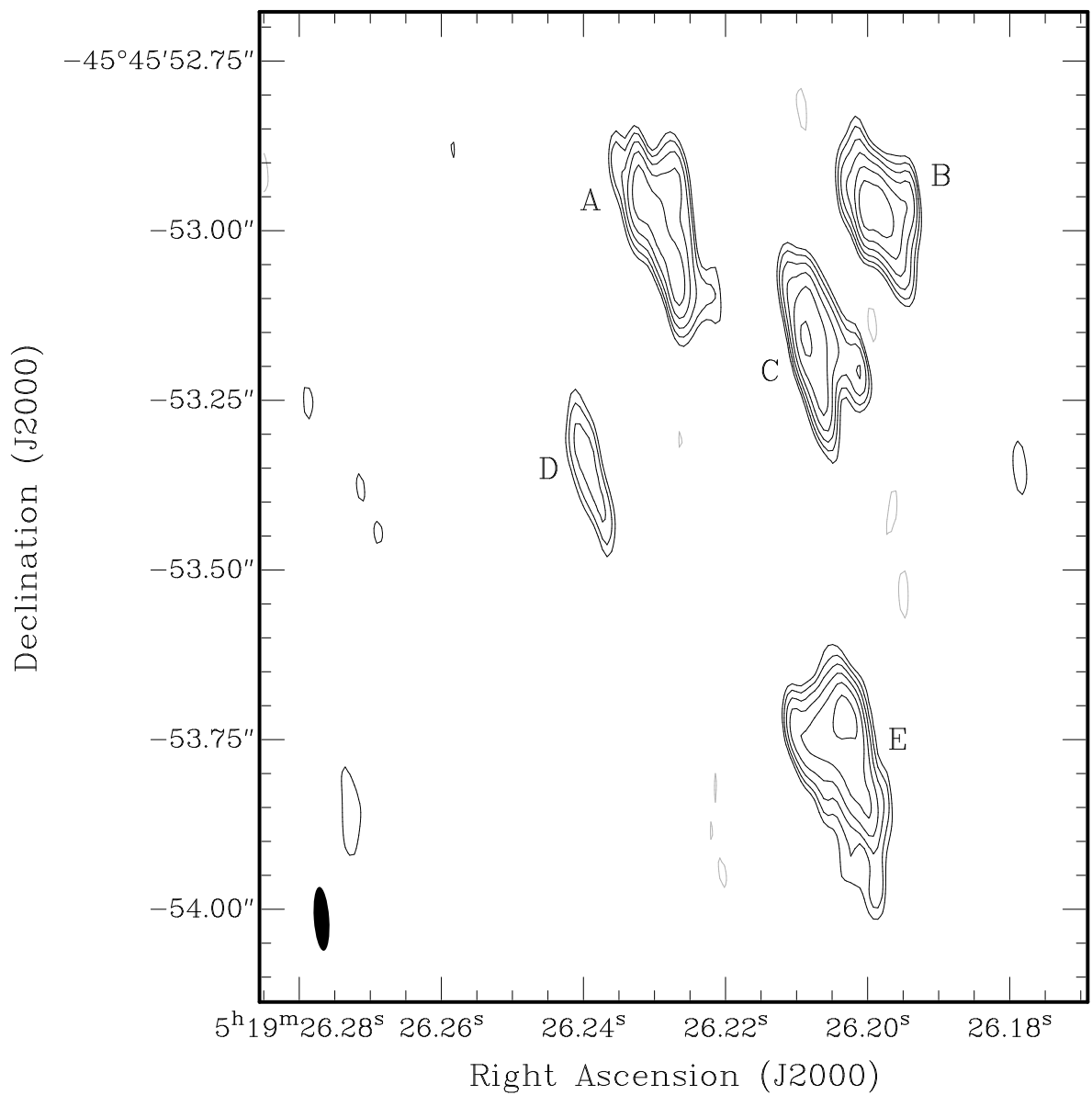}
\caption{VLBA image of the Pictor A hot spot at 18 cm.  The components referred to as A, B, C, D, \& E have their properties tabulated in Table 1.  The image has an RMS noise level of 260 $\mathrm{\mu}$Jy beam$^{-1}$. The peak flux density is 6 mJy beam$^{-1}$ and contours are set at $\pm3$ times the RMS noise level and increase by factors of $\sqrt{2}$. The beam size is approximately $94\times23$ mas at a position angle of $4\arcdeg$.
\label{fig2}}
\end{figure}

\clearpage

\begin{figure}
\center
\epsscale{0.7}
\plotone{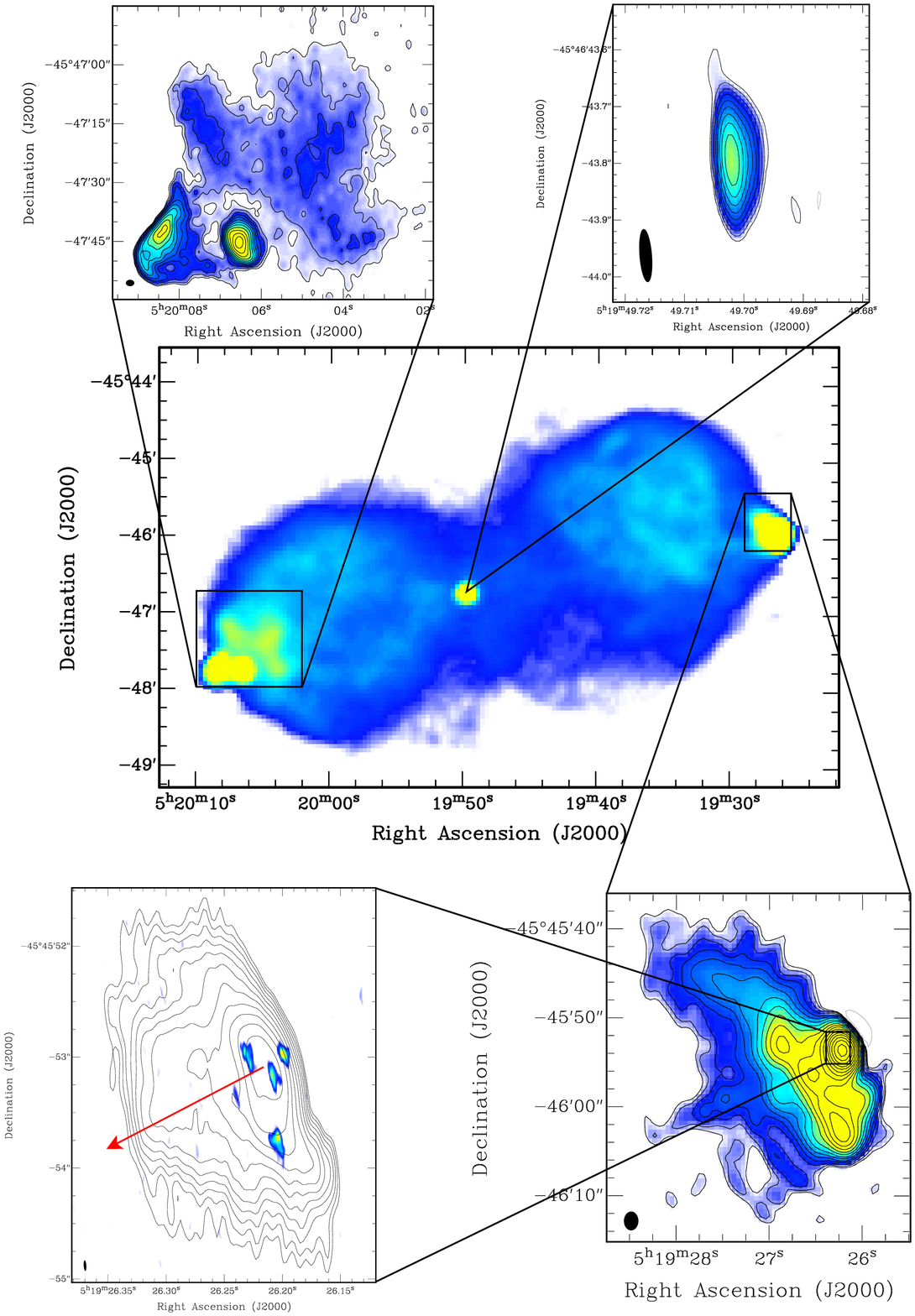}
\caption{Context of the VLBA observations in the overall radio galaxy structure.  The VLA image of \cite{per97} is shown centrally, with zooms into the two hot spots (ATCA data from \cite{len08} and the core (this paper).  A further zoom indicates the high resolution VLBA image of the north-west hot spot, overlaid on the highest resolution VLA image of the hot spot from \cite{per97}.  The red arrow indicates the direction to the nucleus from the location of the peak brightness in the VLA image.\label{fig3}}
\end{figure}

\clearpage

\begin{figure}
\center
\plotone{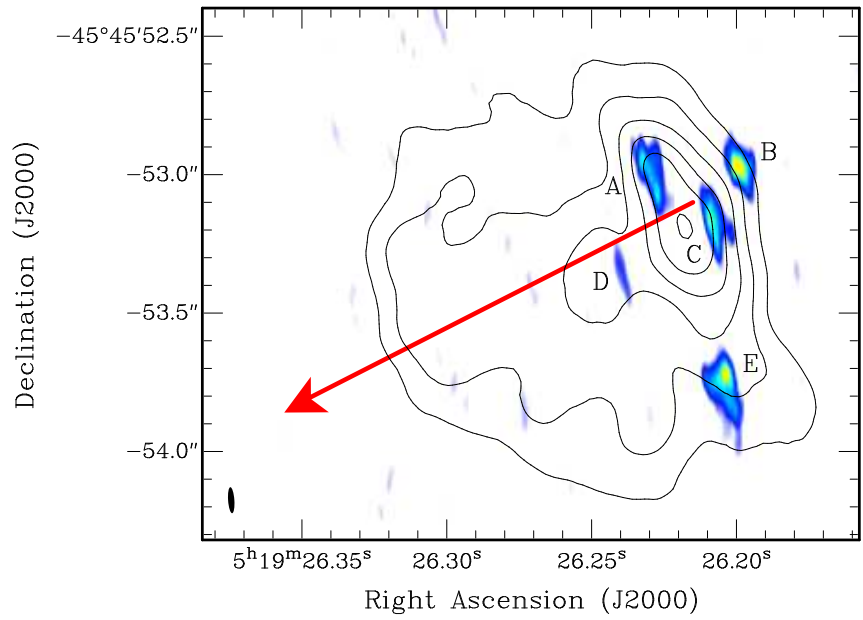}
\caption{HST (contours) and VLBA (gray scale) hot spot overlay.  The red arrow indicates the direction to the nucleus as in Figure 3.  Contours are at 28, 42, 57, 71, \& 85\% of the peak intensity (which has not been calibrated). \label{fig4}}
\end{figure}

\clearpage

\begin{figure}
\center
\plotone{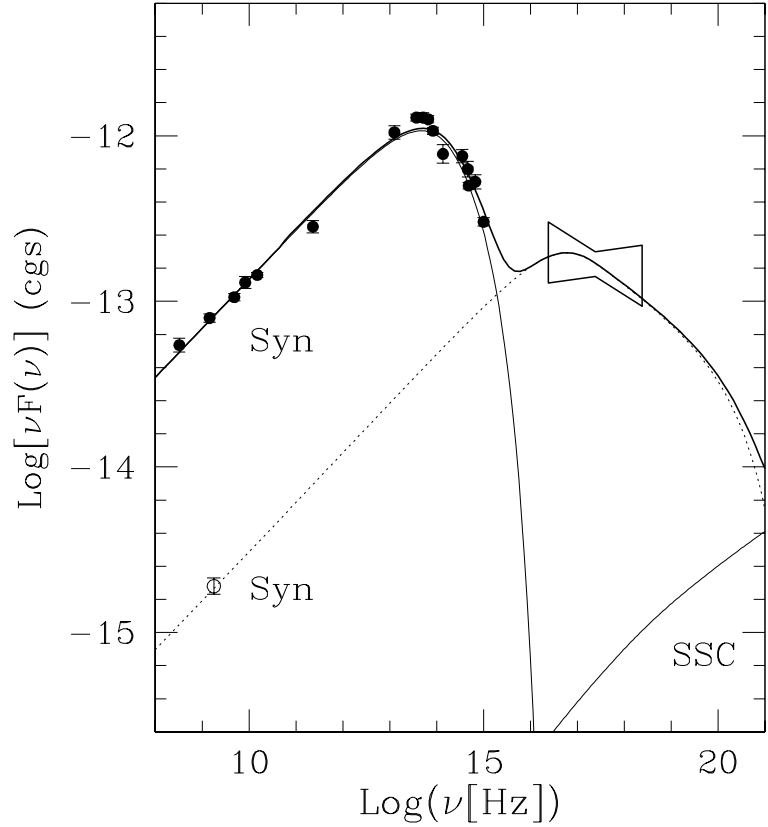}
\caption{The expected synchrotron and inverse Compton spectrum for the Pictor A NW hot spot is compared with the radio, IR, optical and X-ray data.  The synchrotron spectrum is emitted by two components : diffuse hot spot (thin solid line) and pc-scale VLBA regions (dotted line).  The SSC components (thin solid line) is contributed from the hot spot and pc-scale VLBA regions. SSC calculations are reported assuming a reference value of $B=350~\mu$G in both hot spot and compact VLBA regions and adopting the measured size of the components.  The total emission is reported in thick--solid line.
\label{fig5}}
\end{figure}

\clearpage

\begin{figure}
\center
\plotone{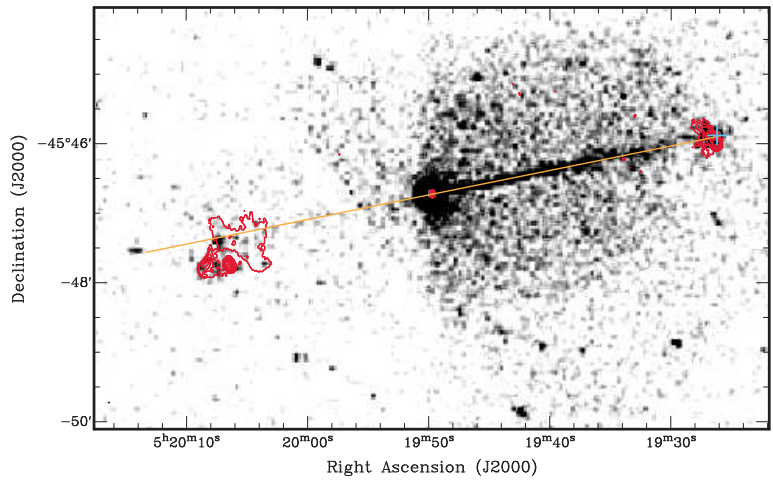}
\caption{Overlay of \emph{Chandra} archival data (gray scale) and ATCA data (contours) from \cite{len08}.  Indicated on the overlay is the jet direction as defined by the jet seen at X-ray wavelengths and the VLBI data of \cite{tin00}, which agree well and both point to the structures seen with the VLBA (indicated by the blue cross).  The jet direction is extrapolated to the south-east, to show the relationship of the south-east hot spot to the expected jet direction, revealing an offset in position angle, as discussed in the text.
\label{fig6}}
\end{figure}

\clearpage

\begin{deluxetable}{lccccccccc}
\tabletypesize{\scriptsize}
\tablecolumns{10}
\tablecaption{Parameters of components in VLBA image of Pictor A NW hot spot}
\tablehead{
   \colhead{}                       &
   \colhead{}                       &
   \colhead{}                       &
   \colhead{}                       &
   \colhead{}                       &
   \multicolumn{3}{c}{FWHM Size}    &
   \multicolumn{2}{c}{FWHM Size}    \\
   \colhead{Component}              &
   \colhead{R.A.}                   &
   \colhead{Decl.}                  &
   \colhead{$S_{P}$}                &
   \colhead{$S_{I}$}                &
   \colhead{major}                  &
   \colhead{minor}                  &
   \colhead{P.A.}                   &
   \colhead{major}                  &
   \colhead{minor}                  \\
   \colhead{}                       &
   \colhead{}                       &
   \colhead{}                       &
   \colhead{(mJy beam$^{-1}$)}      &
   \colhead{(mJy)}                  &
   \colhead{(mas)}                  &
   \colhead{(mas)}                  &
   \colhead{(deg)}                  &
   \colhead{(pc)}                  &
   \colhead{(pc)}
}
\startdata
A   &   05 19 26.2291   &   -45 45 52.993   &   4.3$\pm$0.4   &   28.3$\pm$2.8   &   247   &   62   &   72   &   170   &   42   \\
B   &   05 19 26.1986   &   -45 45 52.967   &   6.3$\pm$0.6   &   25.9$\pm$2.6   &   126   &   67   &   -113   &   87   &   46   \\
C   &   05 19 26.2075   &   -45 45 53.159   &   5.0$\pm$0.5   &   24.9$\pm$2.5   &   185   &   62   &   -100   &   127   &   43   \\
D   &   05 19 26.2392   &   -45 45 53.352   &   1.9$\pm$0.2   &   6.0$\pm$0.6   &   201   &   40   &   -103   &   138   &   28   \\
E   &   05 19 26.2036   &   -45 45 53.751   &   5.7$\pm$0.6   &   36.2$\pm$3.6   &   178   &   85   &   67   &   122   &   58   \\
\enddata
\label{tab:tabcomponents}
\end{deluxetable}

\clearpage

\begin{deluxetable}{ccc}
\tabletypesize{\scriptsize}
\tablecolumns{3}
\tablecaption{Spitzer infrared data for Pictor A north-west hot spot}
\tablehead{
   \colhead{Log$_{10}$($\nu$)}                       &
   \colhead{Log$_{10}$(flux erg/s/Hz)}                       &
   \colhead{Error \%}                       \\
   }
\startdata
 13.10     &      -25.08       &          10 \\
 13.57      &     -25.46       &           5 \\
 13.71      &     -25.60       &           7 \\
 13.82     &      -25.72       &           5 \\
 13.92      &     -25.89       &           5 \\
\enddata
\label{tab:tabIRcomponents}
\end{deluxetable}

\end{document}